\documentclass[showpacs,showkeys]{revtex4}
\usepackage{graphicx}
\usepackage{epstopdf}

\usepackage{amsmath}
\usepackage{amssymb}
\usepackage{color}
\newcommand\gothfamily{\usefont{U}{ygoth}{m}{n}}
\DeclareTextFontCommand{\textgoth}{\gothfamily}

\begin{document}

\title{Canonical conjugated Dirac equation in a curved space}

\author{Vladimir Dzhunushaliev}
\email{vdzhunus@krsu.edu.kg}
\affiliation{Institute for Basic Research,
Eurasian National University,
Astana, 010008, Kazakhstan 
}

\date{\today}

\begin{abstract}
It is shown that the calculation of Dirac operator for the spherical coordinate system with spherical Dirac matrices and using the spin connection formalism is in the contradiction with the definition of standard Dirac operator in the spherical Minkowski coordinate system. It is shown that such contradiction one can avoid by introducing a canonical conjugated covariant derivative for the spinor field. The Dirac equation solution on the Reissner - Nordstr\"om background is obtained. The solution describes a bound state of a charged particle. 
\end{abstract}


\maketitle

\section{Introduction}

Solving of Dirac equation in a curve spacetime is not a simple problem. At the moment we know only the solutions for a spinor field propagating on a curved background (for details, see Ref. \cite{chandrasekhar}). Such solutions do not present a bound state of an electric charge (electron) in a strong gravitational field. The solutions describing a bound state of a charged particle on the background of either a black hole or a gravitational singular point will be physically interesting because they will describe Bohr atom - like configuration created by electron wave function being in a strong gravitational field. In this case the electron undergoes a gravity attraction and Coulomb attraction/repulsion created by gravitational mass and charge. The interesting questions arising in this connection are following:  what force  (gravitational or Coulomb) influences more strongly on a charged particle, the influence of the event horizon for the existence of the solution and so on. 

Let us note that at the moment not too much solutions of both Dirac equation in a curve spacetime and self consistent solutions in Einstein - Dirac gravity are known. In the first case there are well known spinor solutions propagating on a curved background Ref. \cite{chandrasekhar}. In second case there are  cosmological solutions with a spinor field, see Refs. \cite{Saha:2010zza}.

\section{Canonical conjugated Dirac equation in a curved space}

In this section we would like to give arguments in favor that in a curve space we should use a canonical transformed Dirac equation. Let us consider Dirac operator in Minkowski spacetime 
\begin{equation}
	i \gamma^\mu \partial_\mu = 
	\gamma^{\bar 0} \partial_t + \gamma^{\bar 1} \partial_x + 
	\gamma^{\bar 2} \partial_y + \gamma^{\bar 3} \partial_z
\label{1-1}
\end{equation}
where $\gamma^a, a=\bar 0, \bar 1, \bar 2, \bar3$ are standard Dirac matrices 
\begin{equation}
	\gamma^{\bar 0} = \begin{pmatrix}
		1	&	0	&	0	&	0	\\
		0	&	1	&	0	&	0	\\
		0	&	0	&	-1	&	0	\\
		0	&	0	&	0	&	-1	
	\end{pmatrix}, \quad 
	\gamma^{\bar 1} = \begin{pmatrix}
		0	&	0	&	0	&	1	\\
		0	&	0	&	1	&	0	\\
		0	&	-1	&	0	&	0	\\
		-1	&	0	&	0	&	0	
	\end{pmatrix}, \quad 
	\gamma^{\bar 2} = \begin{pmatrix}
		0	&	0	&	0	&	-i	\\
		0	&	0	&	i	&	0	\\
		0	&	i	&	0	&	0	\\
		-i	&	0	&	0	&	0	
	\end{pmatrix}, \quad 
	\gamma^{\bar 3} = \begin{pmatrix}
		0	&	0	&	1	&	0	\\
		0	&	0	&	0	&	-1	\\
		-1	&	0	&	0	&	0	\\
		0	&	1	&	0	&	0	
	\end{pmatrix}.
\label{1-160}
\end{equation}
In order to obtain Dirac operator for the spherical coordinate system we pass from the Cartesian coordinate system to the spherical one using coordinate transformation 
\begin{equation}
	x = r \sin \theta \cos \varphi; \quad 
	y = r \sin \theta \sin \varphi; \quad 
	x = r \cos \theta .
\label{1-180}
\end{equation}
We will obtain the following Dirac operator for the spherical coordinate system 
\begin{equation}
	i \gamma^\mu \partial_\mu = 
		\gamma^{\bar 0} \partial_t + \gamma^{\bar 1} \partial_x + 
		\gamma^{\bar 2} \partial_y + \gamma^{\bar 3} \partial_z
	 = 
		\gamma^{\bar 0} \partial_t + \gamma^{\bar r} \partial_r + 
		\frac{\gamma^{\bar \theta}}{r} \partial_\theta + 
		\frac{\gamma^{\bar \varphi}}{r \sin \theta} \partial_\varphi.
\label{1-170}
\end{equation}
it is easy to show that the Dirac matrices 
$\gamma^{\bar 0, \bar r, \bar \theta, \bar \varphi}$ for the spherical coordinate system in Minkowski space are 
\begin{align}
	\gamma^{\bar 0} &= \begin{pmatrix}
		1	&	0	&	0	&	0	\\
		0	&	1	&	0	&	0	\\
		0	&	0	&	-1	&	0	\\
		0	&	0	&	0	&	-1	
	\end{pmatrix}, \quad 
	&
	\gamma^{\bar r} &= \begin{pmatrix}
		0	&	0	&	\cos \theta	&	\sin \theta e^{-i \phi}	\\
		0	&	0	&	\sin \theta e^{i \phi}	&	-\cos \theta	\\
		-\cos \theta	&	-\sin \theta e^{-i \phi}	& 0	&	0	\\
		-\sin \theta e^{i \phi}	&	\cos \theta	&	0	&	0	
	\end{pmatrix},
\label{1-100} \\
	\gamma^{\bar \theta} &= \begin{pmatrix}
		0	&	0	&	-\sin \theta	&	\cos \theta e^{-i \phi}	\\
		0	&	0	&	\cos \theta e^{i \phi}	&	\sin \theta	\\
		\sin \theta	&	-\cos \theta e^{-i \phi}	& 0	&	0	\\
		-\cos \theta e^{i \phi}	&	-\sin \theta	&	0	&	0	
	\end{pmatrix}, \quad 
	&
	\gamma^{\bar \varphi} &= \begin{pmatrix}
		0	&	0	&	0	&	- i e^{-i \phi}	\\
		0	&	0	&	i e^{i \phi}	&	0	\\
		0	&	i e^{-i \phi}	& 0	&	0	\\
		-i e^{i \phi}	&	0&	0	&	0	
	\end{pmatrix}.
\label{1-120}
\end{align}
The standard Dirac operator $\hat D$ for any (flat or curve) space is 
\begin{equation}
	\hat D = i \gamma^\mu \nabla_\mu 
\label{1-8a}
\end{equation}
where $\gamma^\mu$ are the Dirac matrices in any (flat or curve) space and $\nabla_\mu$ is the covariant derivative for the spinor field $\psi$. In order to determine $\gamma^\mu$ and $\nabla_\mu$ we have to determine tetrads, (for details, see Appendix \ref{app}). 

Now we would like to repeat the result \eqref{1-170} using the spin connection formalism. Let us calculate the Dirac operator for the spherical coordinate system in Minkowski spacetime. The metric is 
\begin{equation}
	ds^2 = dt^2 - dr^2 - r^2 \left(
		d \theta^2 + \sin^2 \theta d \phi^2
	\right)
\label{1-72}
\end{equation}
the tetrads are 
\begin{equation}
\begin{split}
	e^{\bar 0}_{\phantom{0} 0} &= 1, \quad 
	e^{\bar 1}_{\phantom{1} 1} = 1, \quad 
	e^{\bar 2}_{\phantom{2} 2} = r, \quad 
	e^{\bar 3}_{\phantom{3} 3} = r \sin \theta ;
\\
	e_{\bar 0}^{\phantom{0} 0} &= 1, \quad 
	e_{\bar 1}^{\phantom{1} 1} = 1, \quad 
	e_{\bar 2}^{\phantom{2} 2} = \frac{1}{r}, \quad 
	e_{\bar 3}^{\phantom{3} 3} = \frac{1}{r \sin \theta} .
\label{1-201}
\end{split}
\end{equation}
Using the tetrads as \eqref{1-200} we obtain the spin connection 
$\omega_{abc} = e_c^{\phantom{c} \mu} \omega_{ab \mu}$ as 
\begin{equation}
	\omega_{\bar 1 \bar 2 \bar 2} = 
	\omega_{\bar 1 \bar 3 \bar 3} = \frac{1}{r};  \quad 
	\omega_{\bar 2 \bar 3 \bar 3} = \frac{\cot \theta}{r}.
\label{1-210a}
\end{equation}
Now we have the question what kind of Dirac matrices $\gamma^a$ we have to use to obtain Dirac equations ? It is not a trivial question especially for a curve spacetime. For the flat space we can calculate these Dirac matrices using coordinate transformation from old to new coordinate system. In our case we did it and we have Dirac matrices for the spherical coordinate system in  Minkowski spacetime as \eqref{1-100}-\eqref{1-120}. Using \eqref{1-40} we obtain Dirac operator for the spherical coordinate system in Minkowski spacetime 
\begin{equation}
	i \gamma^\mu \nabla_\mu = 
	i \gamma^a e_a^{\phantom{a} \mu} \nabla_\mu = 
	i \left[
		\gamma^0 \partial_t + \gamma^1 \left(
			\partial_r + \frac{1}{r} 
		\right) + 
		\gamma^2 \left(
			\partial_\theta + \frac{\cot \theta}{2}
		\right) + 
		\gamma^3 \partial_\varphi
	\right]
\label{1-240}
\end{equation}
where $\gamma^0 = e_{\bar 0}^{\phantom{0} 0} \gamma^{\bar 0}, 
\gamma^1 = e_{\bar 1}^{\phantom{1} 1} \gamma^{\bar r}, 
\gamma^2 = e_{\bar 2}^{\phantom{2} 2} \gamma^{\bar \theta}, 
\gamma^3 = e_{\bar 3}^{\phantom{3} 3} \gamma^{\bar \varphi}$. But from \eqref{1-170} we see that it should be 
\begin{equation}
	i \gamma^\mu \nabla_\mu = 
	i \left(
		\gamma^0 \partial_t + \gamma^1 \partial_r + 
		\gamma^2 \partial_\theta + 
		\gamma^3 \partial_\varphi
	\right).
\label{1-250}
\end{equation}
Why it happened ? Why we have obtained two different Dirac operators ? In our opinion it happens because we have an arbitrariness in the choice of Dirac matrices $\gamma^a$. Let us note that if we rewrite Dirac operator in \eqref{1-240} as 
\begin{equation}
	i r \sqrt{\sin\theta} \left[
			\gamma^0 \partial_t + \gamma^1 \left(
				\partial_r + \frac{1}{r} 
			\right) + 
			\gamma^2 \left(
				\partial_\theta + \frac{\cot \theta}{2}
			\right) + 
			\gamma^3 \partial_\phi
		\right] \frac{1}{r \sqrt{\sin\theta}} = 
	i \left(
		\gamma^0 \partial_t + \gamma^1 \partial_r + 
		\gamma^2 \partial_\theta + 
		\gamma^3 \partial_\varphi
	\right)
\label{2-30a}
\end{equation}
then both Dirac operators \eqref{1-240} and \eqref{1-250} are the same. 

It allows us to propose the following receipt \textcolor{blue}{\emph{to do Dirac operator to be invariant under the choice of $\gamma^a$}}. We introduce a canonical conjugated covariant derivative in the form 
\begin{equation}
	\tilde \nabla_\mu = \hat S^{-1} \nabla_\mu \hat S 
\label{1-8b}
\end{equation}
where the operator $\hat S$ is chosen in the form to do Dirac equation invariant under the change of Dirac matrices $\gamma^a$. In this case the canonical conjugated Dirac equation is 
\begin{equation}
	i \gamma^\mu \hat S^{-1} \left( 
		\nabla_\mu + i e A_\mu 
	\right) \hat S \psi = m \psi .
\label{1-8}
\end{equation}
where $A_\mu$ is the potential of the Maxwell electromagnetic field. 

\section{Dirac equation in Reissner - Nordstr\"om spacetime}

In this section we will obtain the canonical conjugated Dirac quation on the Reissner - Nordst\"om background. The Reissner-Nordstrom metric is 
\begin{equation}
	ds^2 = \Delta^2(r) dt^2 - \frac{dr^2}{\Delta^2(r)} - r^2 \left(
		d \theta^2 + \sin^2 \theta d \phi^2
	\right)
\label{1-190}
\end{equation}
where $\Delta^2(r) = 1 - \frac{2 M}{r} + \frac{Q^2}{r^2}$. For the Reissner - Nordstr\"om metric we choose the tetrads as 
\begin{equation}
\begin{split}
	e^{\bar 0}_{\phantom{0} 0} &= \Delta, \quad 
	e^{\bar 1}_{\phantom{1} 1} = \frac{1}{\Delta}, \quad 
	e^{\bar 2}_{\phantom{2} 2} = r, \quad 
	e^{\bar 3}_{\phantom{3} 3} = r \sin \theta ;
\\
	e_{\bar 0}^{\phantom{0} 0} &= \frac{1}{\Delta}, \quad 
	e_{\bar 1}^{\phantom{1} 1} = \Delta, \quad 
	e_{\bar 2}^{\phantom{2} 2} = \frac{1}{r}, \quad 
	e_{\bar 3}^{\phantom{3} 3} = \frac{1}{r \sin \theta} .
\label{1-200}
\end{split}
\end{equation}
We would like to emphasize that \textcolor{blue}{\emph{for such choice of the tetrads we will use the Dirac matrices in the form \eqref{1-100}-\eqref{1-120}}}. Using the tetrads as \eqref{1-200} we obtain the spin connection 
$\omega_{abc} = e_c^{\phantom{c} \mu} \omega_{ab \mu}$ as 
\begin{equation}
	\omega_{\bar 0 \bar 1 \bar 0} = \Delta'; \quad 
	\omega_{\bar 1 \bar 2 \bar 2} = 
	\omega_{\bar 1 \bar 3 \bar 3} = \frac{\Delta}{r};  \quad 
	\omega_{\bar 2 \bar 3 \bar 3} = \frac{\cot \theta}{r}.
\label{1-210}
\end{equation}
The operator $\hat S$ we choose as in Eq. \eqref{2-30a} in the form 
\begin{equation}
	\hat S = \frac{1}{\sqrt{\textgoth{e}}}
\label{2-2}
\end{equation}
where $\textgoth{e} = \det e^a_{\phantom{a}\mu}$ is the determinant of the tetrad $e^a_{\phantom{a}\mu}$. 
In this case the canonical transformed Dirac equation will be 
\begin{equation}
	i \gamma^\mu \left[ \textgoth{e} \left( 
		\nabla_\mu  + i e A_\mu 
	\right) \frac{1}{\textgoth{e}} \right] \psi = m \psi .
\label{2-20}
\end{equation}
For the Reissner-Nordstr\"om metric the canonical transformed Dirac equation will be  
\begin{equation}
	i \left\{ 
		r \sqrt{\sin\theta} \left[
			\gamma^0 \left(
				\partial_t + i e \phi 
			\right) + 
			\gamma^1 \left(
				\partial_r + \frac{1}{r} + \frac{\Delta'}{2 \Delta}
			\right) + 
			\gamma^2 \left(
				\partial_\theta + \frac{\cot \theta}{2}
			\right) + 
			\gamma^3 \partial_\phi
		\right] \frac{1}{r \sqrt{\sin\theta}} 
	\right\}\psi = m \psi 
\label{2-30}
\end{equation}
where the spinor $\psi$ we take in the standard form 
\begin{equation}
	\psi = e^{-i \omega t} \begin{pmatrix}
		f(r)	\\
		0	\\
		i g(r) \cos \theta \\
		i g(r) \sin \theta e^{i \varphi}	
	\end{pmatrix}
\label{2-40}
\end{equation}
where $e\phi$ is the potential energy of the charge $e$ in the Reissner-Nordstr\"om electric field. 

That leads to the Dirac operator 
\begin{equation}
	i \left[
		\gamma^0 \left(
				\partial_t + i e \phi 
			\right) + \gamma^1 \left(
			\partial_r + \frac{\Delta'}{2 \Delta}
		\right) + 
		\gamma^2 \partial_\theta + 
		\gamma^3 \partial_\phi
	\right] \psi = m \psi 
\label{1-220}
\end{equation}
where 
\begin{equation}
	\gamma^0 = e_{\bar 0}^{\phantom{0} 0} \gamma^{\bar 0} , \quad 
	\gamma^1 = e_{\bar 1}^{\phantom{1} 1} \gamma^{\bar r} , \quad 
	\gamma^2 = e_{\bar 2}^{\phantom{2} 2} \gamma^{\bar \theta} , \quad 
	\gamma^3 = e_{\bar 3}^{\phantom{3} 3} \gamma^{\bar \varphi} 
\label{1-230}
\end{equation}
\textcolor{blue}{
\emph{we would like once more to emphasize that on the RHS we have to use }
$\gamma^{\bar r, \bar \theta, \bar \varphi}$ \emph{not} 
$\gamma^{\bar 1, \bar 2, \bar 3}$. 
}

After that we have Dirac equation describing a charged particle on the Reissner-Nordstr\"om background 
\begin{eqnarray}
	g' + g \left(
	\frac{2}{r \Delta} + \frac{\Delta'}{2 \Delta} 
	\right)+ \frac{f}{\Delta} \left(
		-m + \frac{\omega - e \phi}{\Delta}
	\right) &=& 0, 
\label{2-50a}\\
	f' + f \frac{\Delta'}{2 \Delta} - \frac{g}{\Delta} \left(
		m + \frac{\omega - e \phi}{\Delta}
	\right) &=& 0
\label{2-60a}
\end{eqnarray}
We scale these equations by the following way 
\begin{equation}
	x = mr; \quad 
	\tilde \omega = \frac{\omega}{m}; \quad 
	\tilde M = \frac{k m M}{c^2}; \quad 
	\tilde e = \sqrt{\frac{k e m}{c^4}}; \quad 
	\tilde Q = \sqrt{\frac{k Q m}{c^4}}; \quad 
	\tilde l_{Pl} = m l_{Pl} = m \sqrt{\frac{\hbar k}{c^3}}
\label{2-70}
\end{equation}
here $k$ is Newton constant; $M,Q$ are the mass and the charge of the Reissner-Nordstr\"om spacetime; $m$ is the mass of the spinor field. After that we have dimensionless equations 
\begin{eqnarray}
	g' + g \left(
	\frac{2}{x \tilde \Delta} + \frac{\tilde \Delta'}{2 \tilde \Delta}
	\right) + 
	\frac{f}{\tilde \Delta} \left(
		- 1 + 
		\frac{\tilde \omega - \frac{\tilde e \tilde Q}{\tilde l^2_{Pl} x}}
		{\tilde \Delta}
	\right) &=& 0, 
\label{2-50}\\
	f' + f \frac{\tilde \Delta'}{2 \tilde \Delta} - 
	\frac{g}{\tilde \Delta} \left(
		1 + 
		\frac{\tilde \omega - \frac{\tilde e \tilde Q}{\tilde l^2_{Pl} x}}
		{\tilde \Delta}
	\right) &=& 0 
\label{2-60}
\end{eqnarray}
where 
$\tilde \Delta = \sqrt{1 - \frac{2 \tilde M}{x} + \frac{\tilde Q^2}{x^2}}$. 

Now we would like to investigate two cases: Reissner-Nordstr\"om black hole with $\tilde M^2 > \tilde Q^2$ and Reissner-Nordstr\"om naked singularity with $\tilde M^2 < \tilde Q^2$. 

\subsection{Reissner-Nordstr\"om black hole with $\tilde M^2 > \tilde Q^2$}

In this case we have two event horizons 
\begin{equation}
	x_{\pm} = \tilde M \pm \sqrt{\tilde M^2 - \tilde Q^2}.
\label{2-1-10}
\end{equation}
Let us consider the behavior of the spinor field on the event horizon $x_+$. From Eq's \eqref{2-50} \eqref{2-60} we see that the spinor field near the  event horizon have to be as following  
\begin{eqnarray}
	g(x) &=& g_1 (x - x_+)^\alpha + \cdots 
\label{2-1-20}\\
	f(x) &=& f_1 (x - x_+)^\alpha + \cdots 
\label{2-1-30}
\end{eqnarray}
The substitution \eqref{2-1-20} \eqref{2-1-30} into \eqref{2-50} \eqref{2-60} gives us the constraint on the parameters 
$\tilde M, \tilde e, \tilde Q, \alpha$
\begin{eqnarray}
	\alpha + \frac{x_+ - \tilde M}{2(x_+ - x_-)} = 0
\label{2-1-40}\\
		\frac{\tilde e \tilde Q}{x_+ \tilde l^2_{Pl}} - \tilde \omega = 0
\label{2-1-50}
\end{eqnarray}
From \eqref{2-1-10} it follows that $\alpha = -1/2$. It means that the condition of the wave function normalization will be destroyed because the corresponding integral  
\begin{equation}
	\int \limits_{r_+}^\infty \left( f^2 + g^2 \right) r^2 dr = \infty 
\label{2-1-60}
\end{equation}
diverges. 

\subsection{Reissner-Nordstr\"om naked singularity with 
$\tilde M^2 < \tilde Q^2$}

Near the singularity $x=0$ the solution of Eq's \eqref{2-50} \eqref{2-60} is given as 
\begin{eqnarray}
	g(x) &=& g_0 x^{1/2} + \cdots ,
\label{2-2-10} \\	
	f(x) &=& f_0 x^{1/2} + \cdots  
\label{2-2-20}
\end{eqnarray}
where $g_0, f_0$ are constants. 

We will solve Eq's \eqref{2-50} \eqref{2-60} numerically. The numerical solution have to be started not from $x=0$ but from the point $x=\delta \ll 1$ because Eq. \eqref{2-50} has the term 
$\frac{\tilde \Delta'}{\tilde \Delta}$.  The boundary conditions are 
\begin{equation}
	g(\delta) = g_0 \delta^{1/2}, \quad f(\delta) = f_0 \delta^{1/2}
\label{2-2-30a}
\end{equation}
where $f_0, g_0$ are arbitrary constants. We consider Eq's \eqref{2-50} \eqref{2-60} as the eigenvalue problem for eigenfunctions $f(x), g(x)$ and eigenvalue $\tilde \omega$. We use shooting method for the computation of the  eigenvalue $\tilde \omega$. 

\begin{figure}[h]
	\fbox{
	\centering
		\includegraphics[width=0.5\linewidth]{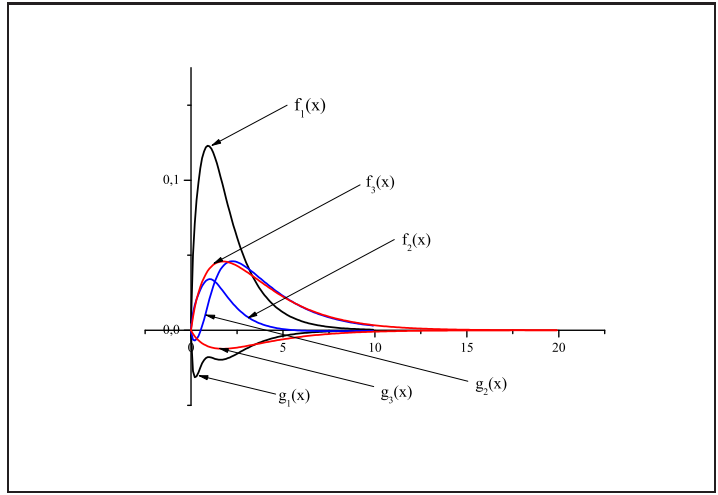}
	}
	\caption{The profiles of $f_1(x), g_1(x)$ for attractive Coulomb 
	interaction in a curve spacetime - black line 
	($\tilde \omega = .76710703, \tilde e = -0.5$); 
	 $f_2(x), g_2(x)$ for repulsive Coulomb interaction in a curve 
	spacetime - \textcolor{blue}{blue line} 
	($\tilde \omega = -0.875205, \tilde e = 0.5$) 
	and $f_3(x), g_3(x)$ attractive Coulomb interaction in Minkowski 
	spacetime - \textcolor{red}{red line} 
	($\tilde \omega = 0.866023716, \tilde e = -0.5$). 
	$f_0 = 0.15$, $g_0=-0.1$, $\tilde M = 0.5$, $\tilde Q = 1.$, 
	$\tilde l_{Pl} = 1.$, $\delta=0.001$.}
	\label{fg}
\end{figure}

In Fig. \ref{fg} the profiles of $f(x)$ and $g(x)$ are presented for attractive $\tilde e<0, \tilde Q>0$ and repulsive $\tilde e>0, \tilde Q>0$ cases. It is necessary to note that for both cases (whether the Coulomb interaction is attractive or repulsive) we have the solution describing a bound state of a charged particle. It happens because the attractive gravitational interaction is much stronger the Coulomb interaction. 

It is interesting to compare these solutions with Bohr atom solution: the corresponding solution is presented in Fig. \ref{fg} also. 

At the infinity Eq's \eqref{2-50} \eqref{2-60} are 
\begin{eqnarray}
	g' + \frac{2}{x} g + f (-1 + \tilde \omega) &\approx& 0, 
\label{2-2-30} \\	
	f' - g (1 + \tilde \omega) &\approx& 0
\label{2-2-40}
\end{eqnarray}
whose asymptotical solution is 
\begin{eqnarray}
	g(x) &\approx& 
	-f_\infty \sqrt{\frac{1 - \tilde \omega}{1 + \tilde \omega}} \; 
	\frac{e^{-x \sqrt{1 - \tilde \omega^2}}}{x^2}, 
\label{2-2-50} \\	
	f(x) &\approx& f_\infty \frac{e^{-x \sqrt{1 - \tilde \omega^2}}}{x^2} 
\label{2-2-60}
\end{eqnarray}
which is the same as for the Bohr atom solution because asymptotically the solution is controlled only by $\tilde \omega = \omega/m$ parameter. 

\section{Conclusions}

Here we have considered relativistic quantum mechanics (Dirac equation) on the Reissner - Nordstr\"om background. Such consideration is very useful because it can shed light on the problem of quantum gravity. The matter is that quantum mechanics in a curved spacetime may have some problems and solving these problems may help to understand the problems arising  the development of quantum gravity theory. The relativistic quantum mechanics on a curved background can be considered as a toy model for the investigation of quantum gravity problems. 

We have shown that there is some problems for using the spin connection formalism by the calculation with the Dirac operator. In order to avoid these problems we have proposed to introduce a canonical conjugated covariant derivative. We think that in this case the Dirac operator will be invariant under the choice of flat Dirac matrices. 

In our knowledge at the moment do not exist solutions describing a bound state of a relativistic charged particle with spin on the Reissner - Nordstr\"om background. Here we have obtained solutions for Dirac equation on the background of Reissner - Nordstr\"om spacetime. We have shown that the solution does exist for the Reissner - Nordstr\"om naked singularity and does not exist for the Reissner - Nordstr\"om black hole. Such solution describes a ground state of an electric charge placed in the Reissner - Nordstr\"om background. We see that this quantum state does exist whether the electric charge is positive or negative. It is explained by the fact that the gravitational attraction is much stronger Coulomb repulsion/attraction. 

\section*{Acknowledgements}

I am grateful to the Research Group Linkage Programme of the Alexander von Humboldt Foundation for the support of this research.

\appendix

\section{Spin connection}
\label{app}

We  introduce tetrads $e^a_{\phantom{a}\mu}$ in the following standard way (for details, see Ref's \cite{poplawski} \cite{ortin})
\begin{equation}
	ds^2 = g_{\mu \nu} dx^\mu dx^\nu = \left( e^a_{\phantom{a}\mu} dx^\mu \right)
	\left( e^b_{\phantom{b}\nu} dx^\nu \right) \eta_{ab}
\label{1-10}
\end{equation}
here $g_{\mu \nu}$ is the metric; $a,b = \bar 0, \bar 1, \bar 2, \bar 3$ are Lorentzian indexes; 
$\mu, \nu = 0,1,2,3$ are world indexes; $\eta_{ab}=\left( +,-,-,- \right)$ is Minkowski metric; the inverse matrix $e_a^{\phantom{a}\mu}$ is defined as 
\begin{equation}
	e^a_{\phantom{a}\mu} e_b^{\phantom{a}\mu} = \delta^a_b , \quad 
	e^a_{\phantom{a}\mu} e_a^{\phantom{a}\nu} = \delta^\nu_\mu .
\label{1-20}
\end{equation}
The standard definition of Dirac equation in a curve space is as follows 
\begin{equation}
	i \gamma^\mu \nabla_\mu \psi = m \psi 
\label{1-30}
\end{equation}
$\nabla_\mu$ is the covariant derivative 
\begin{equation}
	\nabla_\mu \psi = \left( 
		\partial_\mu + \frac{1}{4} \omega_{ab \mu} \gamma^a \gamma^b
	\right) \psi 
\label{1-40}
\end{equation}
where $\omega_{ab \mu} = -\omega_{ba \mu}$ is the spin connection 
\begin{equation}
	\omega_{ab \mu} = - e_a^{\phantom{a}\alpha} e_b^{\phantom{a}\beta} 
	\Delta_{\alpha \beta \mu}
\label{1-50}
\end{equation}
$\Delta_{\alpha \beta \mu}$ are the Ricci coefficients 
\begin{equation}
	\Delta_{\alpha \beta \mu} = 
	e^a_{\phantom{a}\alpha} \Omega_{a \beta \mu} - 
	e^a_{\phantom{a}\beta} \Omega_{a \alpha \mu} - 
	e^a_{\phantom{a}\mu} \Omega_{a \alpha \beta} 
\label{1-60}
\end{equation}
$\Omega_{a \mu \nu}$ are the anholonomy coefficients 
\begin{equation}
	\Omega_{a \mu \nu} = \frac{1}{2} \left(
		e_{a \mu, \nu} - e_{a \nu, \mu}
	\right).
\label{1-70}
\end{equation}
We would like to emphasize that in the definition of spin connection we may use \textcolor{blue}{\emph{any}} Dirac matrices $\gamma^a$ in the tangent space. The Dirac matrices for the curve space with the tetrads $e_a^{\phantom{a}\mu}$ are 
\begin{equation}
	\gamma^\mu = e_a^{\phantom{a}\mu} \gamma^a. 
\label{1-80}
\end{equation}

\end{document}